\documentclass[fleqn]{article}

\usepackage{amsmath,amssymb}
\usepackage{graphicx}
\usepackage[top=0.75in, bottom=0.75in, left=0.75in, right=0.75in]{geometry}
\usepackage[small]{caption}
\usepackage[noblocks]{authblk}
\usepackage{hyperref}

\newcommand{\ie}{\textit{i.e.}}

\begin{document}

\title{{\sf Mass Predictions for Pseudoscalar $(J^{PC}=0^{-+})$ Charmonium and Bottomonium Hybrids in QCD Sum-Rules}}

\author[1]{R.\ Berg}
\author[1]{D.\ Harnett}
\author[2]{R.T.\ Kleiv}
\author[2]{T.G.\ Steele}

\affil[1]{Department of Physics, University of the Fraser Valley,  Abbotsford, BC, V2S 7M8, Canada}
\affil[2]{Department of Physics and
Engineering Physics, University of Saskatchewan, Saskatoon, SK,
S7N 5E2, Canada}

\maketitle

\begin{abstract}
Masses of the pseudoscalar $(J^{PC}=0^{-+})$ charmonium and bottomonium hybrids are determined using QCD Laplace sum-rules. The effects of the dimension-six gluon condensate are included in our analysis and result in a stable sum-rule analysis, whereas previous studies of these states were unable to optimize mass predictions. The pseudoscalar  
charmonium hybrid is predicted to have a mass of approximately $3.8\,{\rm GeV}$ and the corresponding bottomonium prediction is  $10.6\,{\rm GeV}$. Calculating the full correlation function, rather than only the imaginary part, is shown to be necessary for accurate formulation of the sum-rules.  The charmonium hybrid mass prediction is discussed within the context of the $XYZ$ resonances. 
\end{abstract}

\section{Introduction}
Quantum chromodynamics (QCD) seems to allow for hadrons which contain explicit gluonic degrees of freedom \ie, glueballs and hybrids. Despite decades of dedicated effort by experimentalists and theoreticians, no such state has been conclusively identified. Within the context of heavy quarkonia, hybrids can make their presence known in two ways: through $J^{PC}$ quantum numbers that are not permissible for conventional quarkonia (so called exotic hybrids), and through an overpopulation of states with conventional (non-exotic) quantum numbers. In this work we focus on the latter scenario.

A promising area in which to search for such hybrid states has been provided by the recent population boom in the charmonium sector above $D\overline{D}$-threshold~\cite{Olsen,Olsen2,Godfrey:2008nc,Pakhlova,Close2007,MokhtarAPS}. Since 2002, more than a dozen new resonances have been discovered, the so-called $XYZ$ resonances, mainly by the Belle and BaBar collaborations; however, few of these particles fit neatly with a conventional charmonium meson interpretation~\cite{Barnes2005}. Not surprisingly, there has been much speculation that some of the new states lie outside the constituent quark model.

In this article, we analyze pseudoscalar  $(J^{PC}=0^{-+})$ charmonium and bottomonium hybrids using a QCD Laplace sum-rules approach. The pioneering calculations for heavy quark hybrids were handled with a constituent gluon model~\cite{Mandula1977}. Additional computational approaches (relevant to the pseudoscalar sector) include the flux tube model~\cite{Barnes1995} which predicts the lightest charmonium hybrids at~4.1--4.2~GeV as well as lattice QCD~\cite{Perantonis,Liu:2011rn,Liu:2012ze} which yields a quenched prediction of~4.01~GeV and  unquenched predictions of about~4.2~GeV. To our knowledge, the only sum-rules literature concerning heavy quark pseudoscalar hybrids is Refs.~\cite{Govaerts:1985fx,Govaerts:1984hc,Govaerts:1986pp}. As noted therein, the sum-rules that were derived demonstrated instabilities when analyzed. With this paper, we aim to address these instabilities  and update the sum-rule mass prediction by extending previous work~\cite{Govaerts:1985fx,Govaerts:1984hc} to include dimension-six gluon condensate effects.  

In Section~\ref{theSumRules}, we compute the relevant two-point correlation function. We include leading-order perturbative contributions as well as contributions stemming from the dimension-four and dimension-six gluon condensates. Using these results, we then derive the needed Laplace sum-rules.    
In Section~\ref{theAnalysis}, we analyze the sum-rules using the single narrow resonance model and extract ground state mass predictions.
Finally, in Section~\ref{theConclusion}, we comment on our charmonium hybrid results and interpret them within the context of current experimental data. 

\section{Laplace Sum-Rules for the Pseudoscalar Heavy Quark Hybrids}
\label{theSumRules} 
The pseudoscalar ($J^{PC}=0^{-+}$) heavy quark hybrid states can be studied from the following correlation function \cite{Govaerts:1985fx} 
\begin{gather}
\Pi_{\mu\nu}(q)=i\int d^4x \,e^{i q\cdot x}\langle 0\vert T\left[j_\mu(x)j_\nu(0)\right]\vert 0\rangle
\label{basic_corr}
\\
j_\mu=\frac{g}{2}\bar Q\lambda^a\gamma^\nu\tilde G^a_{\mu\nu}Q\,,~\tilde G^a_{\mu\nu}=\frac{1}{2}\epsilon_{\mu\nu\alpha\beta}G^a_{\alpha\beta}\,,
\label{current}
\end{gather} 
where $Q$ denotes a heavy (charm or bottom) quark field.
Within \eqref{basic_corr}, the longitudinal part $\Pi_s$  is of primary interest in this work because it probes the $0^{-+}$ states
\begin{equation}
\Pi_{\mu\nu}(q)=\left(\frac{q_\mu q_\nu}{q^2}-g_{\mu\nu} \right)\Pi_v(q^2)+\frac{q_\mu q_\nu}{q^2}\Pi_s(q^2)~.
\label{corr_tensor}
\end{equation}

The leading-order perturbative and gluon condensate $\langle \alpha G^2\rangle=\langle \alpha G^a_{\mu\nu} G^a_{\mu\nu}\rangle$  contributions to the imaginary part of $\Pi_S$ have previously been  calculated \cite{Govaerts:1985fx,Govaerts:1984hc}, but the resulting sum-rule analysis for the $0^{-+}$ mass was unstable \cite{Govaerts:1984hc,Govaerts:1986pp}.  We extend these results by calculating the  leading-order dimension-six gluon condensate $\langle g^3G^3\rangle=\langle g^3 f_{abc} G^a_{\mu\nu} G^b_{\nu\alpha} G^c_{\alpha\mu}\rangle$ contributions to $\Pi_s$.  As will be seen below, the $\langle g^3G^3\rangle$ contribution is sufficient to stabilize the sum-rule $0^{-+}$ mass prediction. The stabilizing effect of $\langle g^3G^3\rangle$ has also been observed for the sum-rule analysis  of $1^{--}$ heavy quark hybrids \cite{Qiao:2010zh}.

We begin by verifying the leading-order perturbative and $\langle \alpha G^2\rangle$ results \cite{Govaerts:1985fx,Govaerts:1984hc} for $\Pi_s$. Ref.~\cite{Narison:2002pw} advocates the desirability of an independent confirmation of the Ref.~\cite{Govaerts:1985fx,Govaerts:1984hc} results; as such we have calculated the full expression for $\Pi_s$ as opposed to simply reproducing the previously-calculated imaginary part  \cite{Govaerts:1985fx,Govaerts:1984hc}.
 
\begin{figure}[hbt]
\centering
\includegraphics[scale=0.3]{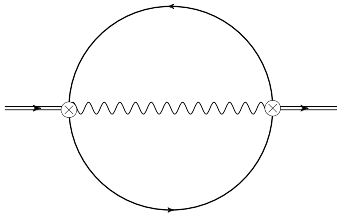}
\caption{Feynman diagram for the leading-order perturbative contribution to $\Pi_s$.  The current is represented by the $\otimes$  symbol.
This and all subsequent Feynman diagrams were created with JaxoDraw \cite{Binosi:2003yf}.
}
\label{pert_fig}
\end{figure}

The leading-order perturbative contribution to $\Pi_s$ is represented in Fig.~\ref{pert_fig}.  We use the Tarcer  \cite{Mertig:1998vk} implementation of loop-integral recurrence relations and tensor structures \cite{Tarasov:1997kx,Tarasov:1996br}  to express $\Pi_s$ in terms of the small set of  basic integrals given in Refs.~\cite{Boos:1990rg,Davydychev:1990cq,Broadhurst:1993mw}. In $D=4+2\epsilon$ dimensions in the $\overline{{\rm MS}}$ scheme, the perturbative result is
\begin{equation}
\begin{split}
\Pi_s^{{\rm pert}}(q^2)=\frac{m^6\alpha}{5400\pi^3}
&\Biggl[
180(z-1)\left(4z^2-21z+10 \right) \phantom{}_3F_2\left(1,1,1;3/2, 3;z\right)
\Biggr.\\
&\Biggl.\qquad
+20z\left(8z^3+8z^2+29z-10  \right)
 \phantom{}_3F_2\left(1, 1, 2; 5/2, 4;z\right)
\Biggr] \,, \quad z=\frac{q^2}{4\,m^2} \,,
\end{split}
\label{Pi_pert}
\end{equation}
where $m$ is the quark mass, and non-physical terms corresponding to dispersion relation subtraction constants have been omitted. The quantities $\alpha$ and $m$ are implicitly evaluated at the renormalization scale $\mu$ in the $\overline{{\rm MS}}$ scheme. Standard conventions for generalized hypergeometric functions have been used (see for example Ref.~\cite{Bateman:1953}). The analytic structure of $\Pi_s$ (\ie, a branch starting at $q^2=4m^2$) is clearly evident from the hypergeometric functions. Analytic continuation of the hypergeometric functions in \eqref{Pi_pert} gives
\begin{equation}
\begin{split}
{\rm Im}\Pi_s^{\rm pert}(q^2)=
\frac{\alpha m^6}{120\pi^2z^2}
&\Biggl(
\sqrt{z-1} \sqrt{z} \left(30-115 z+166 z^2+8 z^3+16 z^4\right)
\Biggr.\\
&\Biggl.\qquad
-15 \left(-2+9 z-16 z^2+16 z^3\right) \log\left[\sqrt{z-1}+\sqrt{z}\,\right]
\Biggr)\,,\quad z>1\,.
\end{split}
\label{Im_Pi_pert}
\end{equation}
Calculating the integral representations for  ${\rm Im}\Pi^{\rm pert}_s$ given in \cite{Govaerts:1985fx,Govaerts:1984hc} we find complete agreement with \eqref{Im_Pi_pert}.

\begin{figure}[hbt]
\centering
\includegraphics[scale=0.3]{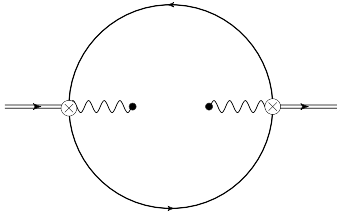}
\caption{Feynman diagram for the leading-order $\langle \alpha G^2\rangle$  contribution to $\Pi_s$.
}
\label{GG_fig}
\end{figure}

The leading order $\langle \alpha G^2\rangle$ contribution to $\Pi_s$ is represented in Fig.~\ref{GG_fig}.  We choose to calculate this contribution using fixed-point gauge methods (see, e.g., Ref.~\cite{Elias:1987ac} for examples applying these methods), which have been proven to be equivalent to  plane-wave techniques for correlation functions of gauge-invariant currents \cite{Bagan:1992tg}.\footnote{Implementation of fixed-point gauge methods is trivial because the field strength appears in the current \eqref{current}. }  Using the same loop-calculation methods as for the perturbative contributions, the  $\langle \alpha G^2\rangle$ result is
\begin{equation}
\Pi^{\rm GG}_s(q^2)=\frac{ \langle \alpha G^2\rangle}{36\pi} m^2z(4z+2) \phantom{}_2F_1\left(1, 1; 5/2;z\right)
\,,
\label{Pi_GG}
\end{equation}
where non-physical terms corresponding to dispersion relation subtraction constants have been omitted. 
The imaginary part of \eqref{Pi_GG}
\begin{equation}
{\rm Im}\Pi^{\rm GG}_s(q^2)=\frac{ m^2\langle \alpha G^2\rangle}{12}\left(1+2z\right)\frac{\sqrt{z-1}}{\sqrt{z}}\,,\quad z>1
\label{Im_Pi_GG}
\end{equation}
again agrees with the explicit result of Refs.~\cite{Govaerts:1985fx,Govaerts:1984hc}.

\begin{figure}[hbt]
\centering
\includegraphics[scale=0.6]{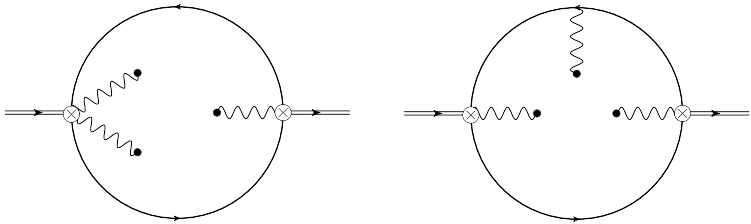}
\caption{Feynman diagram for the leading-order $\langle g^3 G^3\rangle$  contribution to $\Pi_s$. Additional diagrams related by symmetry are not shown.
}
\label{GGG_fig}
\end{figure}

The dimension-six gluon condensate contributions represented by the diagrams in Fig.~\ref{GGG_fig} were not calculated in Refs.~\cite{Govaerts:1985fx,Govaerts:1984hc}.  Using the fixed-point gauge and  loop calculation techniques described above, we find
\begin{equation}
\begin{split}
\Pi_s^{\rm GGG}(q^2)=&\frac{\langle g^3G^3\rangle}{384\pi^2 (z-1)^2}
\left[
4z^2(z-1)-\left(4z^3-6z^2+2z-1  \right)
\right] \phantom{}_2F_1\left(1, 1; 5/2;z\right)
\\&+
\frac{\langle g^3G^3\rangle}{384\pi^2 (z-1)^2}
\left[31z^2-50z+16+(z-1)(9-21z)\right]\,.
\end{split}
\label{Pi_GGG}
\end{equation}
The corresponding imaginary part of \eqref{Pi_GGG} is
\begin{equation}
{\rm Im}\Pi_s^{\rm GGG}(q^2)=\frac{\langle g^3G^3\rangle}{256\pi z(z-1)^2}\frac{\sqrt{z-1}}{\sqrt{z}}\left[
4z^2(z-1)-\left(4z^3-6z^2+2z-1  \right)
\right] \,,\quad z>1\,.
\label{Im_Pi_GGG}
\end{equation} 
At this stage, we note that only the hypergeometric terms contribute to the imaginary part, but as will be shown below, the remaining terms do contribute to the QCD Laplace sum-rules because of the single and double poles at $z=1$.  Thus if only the imaginary parts are calculated (as in Refs.~\cite{Govaerts:1985fx,Govaerts:1984hc}), there exists the possibility that the resulting sum-rule will be inaccurate.  

The dispersion relation appropriate to the 
asymptotic (perturbative) behaviour of $\Pi_s$ is
\begin{equation}
\Pi_s\left(Q^2\right)=\Pi_s(0)+Q^2\Pi_s'(0)+\frac{1}{2}Q^4\Pi_s''(0)+\frac{1}{6}Q^6\Pi_s'''(0)
+Q^8\frac{1}{\pi}\int\limits_{t_0}^\infty
dt\,\frac{ \rho(t)}{t^4\left(t+Q^2\right)} \quad .
\label{disp_rel}
\end{equation}
where $Q^2=-q^2$ is the Euclidean momentum and $\rho(t)$ is the hadronic spectral function with physical threshold $t_0$.  Note that the high-energy behaviour of \eqref{Im_Pi_pert} ensures convergence of the integral.
Direct application of the dispersion relation is not possible because $\Pi_s$ contains field theoretical divergences that are polynomials in $Q^2$ and the associated subtraction constants on the right-hand side of \eqref{disp_rel} are unknown. A related problem is the contribution of excited states and the QCD continuum to the integral of $\rho(t)$ in \eqref{disp_rel}. Enhancement of the lowest-lying resonance contribution in hadronic systems requires greater high-energy suppression of this integral.

The established technique for dealing with these issues is the Laplace sum-rules \cite{Shifman:1978bx,Shifman:1978by}. A family of Laplace sum-rules can be obtained from the dispersion relation \eqref{disp_rel} through the Borel transform operator $\hat B$
\begin{equation}
\hat B\equiv 
\lim_{\stackrel{N,~Q^2\rightarrow \infty}{N/Q^2\equiv \tau}}
\frac{\left(-Q^2\right)^N}{\Gamma(N)}\left(\frac{d}{dQ^2}\right)^N\,,
\label{borel}
\end{equation}  
which has the following useful properties in the construction of the Laplace sum-rules:
\begin{gather}
\hat B\left[a_0+a_1Q^2+\ldots a_m Q^{2m}\right]=0\quad,\quad (m~{\rm finite})
\label{borel_poly}\\
\hat B \left[ \frac{Q^{2n}}{t+Q^2}\right]=\tau \left(-1\right)^nt^ne^{-t\tau}  \quad,\quad n=0,~1,~2,\ldots ~ 
(n~{\rm finite})\,.
\label{borel_exp}
\end{gather}
The Borel transform is related to the inverse Laplace transform via \cite{Bertlmann:1984ih}
\begin{gather}
f\left(Q^2\right)=\int\limits_0^\infty d\tau F(\tau) e^{-Q^2\tau}\equiv{\cal L}\left[ F(\tau)\right]
\quad \Longrightarrow\quad \frac{1}{\tau}\hat B\left[ f\left(Q^2\right)\right]
=F(\tau)={\cal L}^{-1}
\left[ f\left(Q^2\right)\right]
\label{borel_laplace}\\
{\cal L}^{-1}
\left[ f\left(Q^2\right)\right]=\frac{1}{2\pi i}\int\limits_{b-i\infty}^{b+i\infty}
f\left(Q^2\right) e^{Q^2\tau}\,dQ^2
\label{inv_lap_def}
\end{gather}
where $f\left(Q^2\right)$ is analytic to the right of the integration contour  in the complex plane.

The theoretically-determined quantity
\begin{equation}
{\cal L}_k(\tau)\equiv\frac{1}{\tau}\hat B\left[\left(-1\right)^k Q^{2k}\Pi_s\left(Q^2\right)\right]\quad ,  
\label{laplace}
\end{equation}
leads to the following family of Laplace sum-rules, after application of  $\hat B$ to the dispersion relation \eqref{disp_rel} weighted by the appropriate power of $Q^2$:
\begin{equation}
{\cal L}_{k}(\tau)=\frac{1}{\pi}\int\limits_{t_0}^\infty
dt\,t^k e^{-t\tau}\rho(t)\quad ,\quad k\ge 0\,.
\label{lap_gen_k}
\end{equation}
On the right-hand side of~(\ref{lap_gen_k}), we impose the standard resonance plus continuum model
\begin{equation}
  \rho(t)=\rho^{{\rm had}}(t)+\theta\left(t-s_0\right){\rm Im}\Pi^{\rm QCD}(t)
\label{respcont}
\end{equation}
where $s_0$ represents the onset of the QCD continuum. The resulting continuum contribution 
\begin{equation}\label{continuum}
   {\cal L}_k^{{\rm cont}} (\hat{s},\tau,s_0) =   \frac{1}{\pi}  \int_{s_0}^{\infty} t^k
   \exp \left[-t\tau  \right]  {\rm Im} \Pi^{\rm QCD}(t)\; dt
\end{equation}
is then moved to the left-hand side of~(\ref{lap_gen_k}). The total QCD contribution
\begin{equation}
  {\cal L}_k^{\rm QCD}\left(\tau,s_0\right) \equiv {\cal L}_k\left(\tau\right) -  {\cal L}_k^{{\rm cont}} \left(\tau,s_0\right)
\label{blah}
\end{equation}
is then related to the hadronic spectral function
\begin{equation}
 {\cal L}_{k}^{\rm QCD}\left(\tau,s_0\right)  = \frac{1}{\pi}\int_{t_0}^{\infty} t^k
   \exp\left[ -t\tau\right] \rho^{\rm had}(t)\; dt
   ~.
\label{final_gauss}
\end{equation}

We also note that the tensor decomposition in \eqref{corr_tensor} could have been chosen without the overall factor of $1/q^2$, as is done for axial-vector and vector correlators (for examples of each see Ref.~\cite{Narison:2002pw}). If this convention had been used, the perturbative calculation would have a $1/q^2\epsilon^2$ divergence, which must be eliminated by additional weights of $Q^2$ in the Laplace sum-rule \eqref{laplace}. In other words, knowledge of the divergence structure, which is not revealed in the imaginary part, places a bound on the lowest-possible weight $k$ in the Laplace sum-rule \eqref{laplace}.

The Laplace sum-rule can now be calculated using the methods described above (see, e.g., Ref.~\cite{Harnett:2000xz} for detailed examples of applying inverse Laplace transform techniques), leading to the following results
\begin{gather}
\begin{split}
{\cal L}_0^{\rm QCD}\left(\tau,s_0\right)=&\frac{4m^2}{\pi}\int_1^{s_0/4m^2} \left[{\rm Im}\Pi_s^{\rm pert}\left(4m^2 x\right)+{\rm Im}\Pi_s^{\rm GG}\left(4m^2 x\right)\right]\exp{\left(-4m^2\tau x\right)\,dx}
\\
&+\lim_{\eta\to 0^+}\left[\frac{4m^2}{\pi}\int_{1+\eta}^{s_0/4m^2} {\rm Im}\Pi_s^{\rm GGG}(4m^2 x)\exp{\left(-4m^2\tau x\right)\,dx}
-\frac{4m^2\langle g^3G^3\rangle}{128\pi^2\sqrt{\eta}}\exp{(-4m^2\tau)}
\right]
\,,
\end{split}
\label{L_0}
\\
{\cal L}_1^{\rm QCD}\left(\tau,s_0\right)=-\frac{\partial}{\partial\tau}{\cal L}_0^{\rm QCD}\left(\tau,s_0\right)\,.
\label{L_1}
\end{gather}
Several clarifying remarks on Eqs.~(\ref{L_0}, \ref{L_1}) are needed.  First, the mass $m$ and strong coupling $\alpha$ are implicitly $\overline{\rm MS}$-scheme running quantities evaluated at a scale $\mu$.  However, one often implements renormalization-group improvement by setting $\mu=1/\sqrt{\tau}$ \cite{Narison:1981ts}, which must be done after calculating the $\tau$ partial derivative in \eqref{L_1}.  Second, the $\eta$ limiting procedure naturally originates from the inverse Laplace transform approach applied to the full result \eqref{Pi_GGG}, and ensures cancellation of integration divergences arising from the $1/(z-1)$ poles of \eqref{Im_Pi_GGG}. 
Thus, as mentioned earlier, the last term in \eqref{L_0} requires knowledge of the full $\langle g^3G^3\rangle$ contributions and cannot be obtained solely from ${\rm Im}\Pi_s^{GGG}$. 

\section{Analysis: Mass Predictions for the Pseudoscalar Heavy Quark Hybrids}  
\label{theAnalysis}
We analyze the QCD Laplace sum-rules using a single narrow narrow resonance model
\begin{equation}
 \frac{1}{\pi}\rho^{\rm had}(t)=f^2\delta\left(t-M^2\right)\,.
 \label{narrow_res}
\end{equation}
In this approximation, the sum-rules become
\begin{equation}
{\cal L}_k^{\rm QCD}\left(\tau,s_0\right)=f^2 M^{2k}\exp{\left(-M^2\tau\right)}\,,
\label{narrow_sr}
\end{equation}
and the $0^{-+}$ hybrid mass $M$ is given by the ratio of the first two Laplace sum-rules
\begin{equation}
M^2=\frac{{\cal L}_1^{\rm QCD}\left(\tau,s_0\right)}{{\cal L}_0^{\rm QCD}\left(\tau,s_0\right)}\,.
\label{ratio}
\end{equation}
Using fairly general arguments, one can demonstrate that the narrow-width mass estimate would overestimate the actual mass when resonance width effects are included \cite{Elias:1998bq}.  Furthermore, the $s_0\to\infty$ limit provides  an upper bound on the ratio \eqref{ratio}, permitting a very robust upper bound on the ground state mass prediction that is essentially independent of the QCD continuum approximation and resonance model.

Before proceeding with the detailed analysis, the QCD parameters will be specified.  It is easy to see that the quark mass $m$ sets the basic scale of the mass prediction, so it is the most crucial parameter in our analysis.  We have chosen to focus on sum-rule estimates of quark masses as they would provide the greatest possibility of a self-consistent  prediction for the hybrid mass. In particular, the following values encompass the $\overline {\rm MS}$ quark masses of Refs.~\cite{Chetyrkin:2009fv,Narison:2011rn,Narison:2010cg,Kuhn:2007vp}
\begin{gather}
m_c\left(\mu=m_c\right)=\overline m_c=\left(1.28\pm 0.02\right)\,{\rm GeV}\,,
\label{mc_mass}
\\
m_b\left(\mu=m_b\right)=\overline m_b=\left(4.17\pm 0.02\right)\,{\rm GeV}~.
\label{mb_mass}
\end{gather}
These values are within the Particle Data Group's recommended ranges \cite{pdg}.

Since our calculation is leading order, one-loop $\overline{\rm MS}$ expressions for the renormalization group evolution of the strong coupling and quark masses are appropriate.
For the hybrid charmonium analysis, the strong coupling is best determined by evolution from the $\tau$ mass, and in the hybrid bottomonium case by evolution from the $Z$ mass:
\begin{gather}
\alpha(\mu)=\frac{\alpha\left(M_\tau\right)}{1+\frac{25\alpha\left(M_\tau\right)}{12\pi}\log{\left(\frac{\mu^2}{M_\tau^2}\right)}}
\,,~\alpha\left(M_\tau\right)=0.33\,;
\\
\alpha(\mu)=\frac{\alpha\left(M_Z\right)}{1+\frac{23\alpha\left(M_Z\right)}{12\pi}\log{\left(\frac{\mu^2}{M_Z^2}\right)}}
\,,~\alpha\left(M_Z\right)=0.118\,.
\end{gather} 
The numerical values of $\alpha\left(M_\tau\right)$ and $\alpha\left(M_Z\right)$ are based on the determinations of \cite{Bethke:2009jm}, and we use Particle Data Group values of the $\tau$ and $Z$ masses \cite{pdg}.
 The scale dependence of the $\overline{\rm MS}$ masses can then be expressed to the same leading-order as
\begin{gather}
m_c(\mu)=\overline m_c\left(\frac{\alpha(\mu)}{\alpha\left(\overline m_c\right)}\right)^{12/25}\,,
\\
m_b(\mu)=\overline m_b\left(\frac{\alpha(\mu)}{\alpha\left(\overline m_b\right)}\right)^{12/23}\,.
\end{gather}
We set $\mu=1/\sqrt{\tau}$ in our sum-rule analysis \cite{Narison:1981ts}.

For the QCD condensates, we use the following determinations of the QCD condensates from  heavy-quark systems \cite{Narison:2010cg}:
\begin{gather}
\langle g^3G^3\rangle=\left(8.2\pm 1.0\right){\rm GeV^2}\langle \alpha G^2\rangle\,
\label{GGG_value}
\\
\langle \alpha G^2\rangle=\left(7.5\pm 2.0\right)\times 10^{-2}\,{\rm GeV^4}\,.
\label{GG_value}
\end{gather}
For the $\langle g^3G^3\rangle$ contributions to \eqref{L_0}  we find that $\eta=10^{-4}$ is sufficient to evaluate the limit.

\begin{figure}[hbt]
\centering
\includegraphics[scale=0.8]{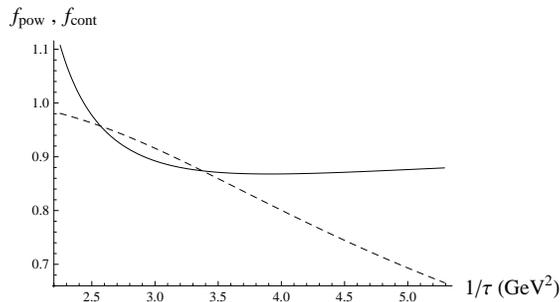}
\caption{The quantities $f_{pow}\left( \tau,s_0 \right)$ (solid line) and $f_{cont}\left( \tau,s_0 \right)$ (dashed line) for hybrid charmonium are shown as a function of the Borel scale $1/\tau$ for the optimized value $s_0=23.0\,{\rm GeV^2}$. Central values of the QCD parameters have been used.
}
\label{charm_window}
\end{figure}

We begin with the analysis of hybrid charmonium.   We first establish a window for the Borel parameter $\tau$ for which  the sum-rule analysis is considered reliable.  Following \cite{Shifman:1978by}, we define the quantities
\begin{gather}
f_{cont}\left( \tau,s_0 \right)=\frac{ {\cal L}_1^{\rm QCD}\left(\tau,s_0\right)/{\cal L}_0^{\rm QCD}\left(\tau,s_0\right) }{{\cal L}_1^{\rm QCD}\left(\tau,\infty\right)/{\cal L}_0^{\rm QCD}\left(\tau,\infty\right) }
\label{f_cont}
\\
f_{pow}\left( \tau,s_0 \right)=\frac{ {\cal L}_1^{\rm QCD}\left(\tau,s_0\right)/{\cal L}_0^{\rm QCD}\left(\tau,s_0\right) }{{\cal L}_1^{\rm pert}\left(\tau,s_0\right)/{\cal L}_0^{\rm  pert}\left(\tau,s_0\right) }\,,
\label{f_pow}
\end{gather}
where ${\cal L}_k^{\rm pert}$ includes only the perturbative corrections arising from \eqref{L_0}.  The ratio \eqref{f_cont} represents the relative importance of the continuum to the sum-rule ratio \eqref{ratio} while \eqref{f_pow} represents the relative importance of non-perturbative (power-law) effects.  Consistent with \cite{Shifman:1978by},  we define the window of sum-rule validity by $f_{cont}>0.7$ (i.e., the continuum contribution does not exceed 30\%) and $0.9< f_{pow}<1.1$ (i.e., the non-perturbative contributions do not exceed 10\%).  
Figure \ref{charm_window} show the resulting constraints on the Borel parameter $\tau$ for the optimum value of $s_0$ to be discussed below.  The resulting region of validity $2.6\,{\rm GeV^2} < 1/\tau <4.8\,{\rm GeV^2}$ is comparable to the window established for the $1^{--}$ charmonium hybrid \cite{Qiao:2010zh}.   If we change to the pole scheme for the charm quark  (with a pole mass $m_c^{pole}=1.71\,{\rm GeV}$ \cite{pdg}), the sum-rule window diminishes considerably and thus the sum-rule is less reliable than in the $\overline{{\rm MS}}$ quark mass scheme, consistent with the findings of \cite{Jamin:2001fw}.  

\begin{figure}[hbt]
\centering
\includegraphics[scale=0.8]{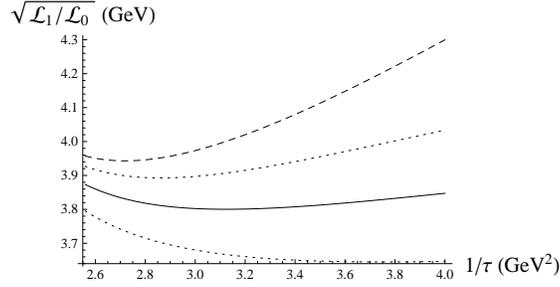}
\caption{The ratio ${\cal L}_1^{\rm QCD}\left(\tau,s_0\right)/{\cal L}_0^{\rm QCD}\left(\tau,s_0\right)$
for hybrid charmonium is shown  as a function of the Borel scale $1/\tau$ for
the optimized value $s_0=23\,{\rm GeV^2}$ (solid curve).  
For comparison the ratio is also shown for $s_0=28\,{\rm GeV^2}$ (upper dotted curve) and $s_0=19\,{\rm GeV^2}$ (lower dotted curve). 
The uppermost dashed curve represents the $s_0\to\infty$ limit corresponding to the bound $M<3.96\,{\rm GeV}$.
Central values of the QCD parameters have been used.}
\label{charm_opt}
\end{figure}

The optimized value of $s_0$ and mass prediction is obtained by finding the minimum value  $s_0=19\,{\rm GeV^2}$ for which the ratio \eqref{ratio} stabilizes (in this case, a minimum) at a $\tau$ value within the $s_0$-dependent region of validity.  We thereby establish a region $s_0>19\,{\rm GeV^2}$
and $1.6\, {\rm GeV}<1/\sqrt{\tau}<2.0\,{\rm GeV}$  for locating  an optimized prediction.  We then search for the optimized mass prediction  $M$ and $s_0$ that minimize the quantity\footnote{A fit based on the quantity  $\frac{1}{M^2}\frac{{\cal L}_1^{\rm QCD}\left(\tau_j,s_0\right)}{{\cal L}_0^{\rm QCD}\left(\tau_j,s_0\right)}$ leads to virtually identical optimizations.  }
\begin{equation}
\chi^2\left(s_0\right)=\sum_j \left( \frac{1}{M}\sqrt{\frac{{\cal L}_1^{\rm QCD}\left(\tau_j,s_0\right)}{{\cal L}_0^{\rm QCD}\left(\tau_j,s_0\right)}}-1 \right)^2\,,
\end{equation}
where $1.6\, {\rm GeV}<1/\sqrt{\tau_j}<2.0\,{\rm GeV}$.  This procedure results in  $s_0=23.0\,{\rm GeV^2}$ and  the predicted charmonium hybrid mass $3.82\,{\rm GeV}$.   In Fig.~\ref{charm_opt} we show the optimized ratio for $s_0=23\,{\rm GeV^2}$  in addition to larger and smaller values, including the $s_0\to\infty$ limit used for obtaining mass bounds.  

Uncertainties in the mass prediction resulting from the QCD input parameters are dominated by  variations of the charm quark mass   \eqref{mc_mass} and $\langle g^3G\rangle$ \eqref{GGG_value}, while $\langle \alpha G^2\rangle$ variations \eqref{GG_value} are relatively stable.  The analysis is also stable under an alternative choice of renormalization scale ($\mu=m_c^{pole}=1.71\,{\rm GeV}$).
 Adding the uncertainties in quadrature, we find the predicted value of the charmonium hybrid mass to be $M=(3.82\pm 0.13)\,{\rm GeV}$.  The influence of $\langle g^3G\rangle$ on the mass prediction corroborates our key observation that the dimension-six condensate effects are essential for stabilizing the mass prediction.
The basic scales of our analysis align well with Ref.~\cite{Qiao:2010zh} which also included effects of $\langle g^3G^3\rangle$ to find a $1^{--}$ charmonium hybrid mass of approximately $4.4\,{\rm GeV}$ for $s_0\approx 26\,{\rm GeV^2}$. By contrast, Ref.~\cite{Govaerts:1985fx}  was not able to obtain optimized mass predictions for $0^{-+}$ and $1^{--}$ hybrid charmonium, so we speculate that the dimension-six condensate $\langle g^3G^3\rangle$ is a necessary component of sum-rule analyses for heavy quark hybrids.  

For hybrid bottomonium a simple scaling behaviour in moving from the  hybrid charmonium to bottomonium systems will not occur because in addition to a function of $q^2/m^2$ that would lead to scaling behaviour,  there are differing pre-factors of the quark mass  for each contribution [see Eqs.~(\ref{Pi_pert},\ref{Pi_GG},\ref{Pi_GGG})]. This will lead to differing weights of the various perturbative and non-perturbative contributions, and hence there are intrinsic field-theoretical differences between hybrid charmonium and bottomonium systems.  

\begin{figure}[hbt]
\centering
\includegraphics[scale=0.8]{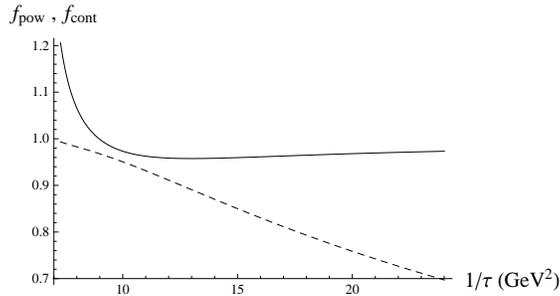}
\caption{The quantities $f_{pow}\left( \tau,s_0 \right)$ (solid line) and $f_{cont}\left( \tau,s_0 \right)$ (dashed line) for hybrid bottomonium are shown as a function of the Borel scale $1/\tau$ for the optimized value $s_0=140\,{\rm GeV^2}$. Central values of the QCD parameters have been used.
}
\label{bottom_window}
\end{figure}

\begin{figure}[hbt]
\centering
\includegraphics[scale=0.8]{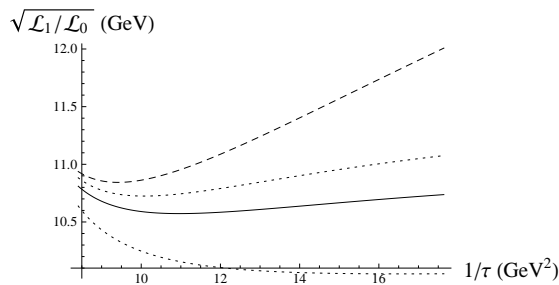}
\caption{The ratio ${\cal L}_1^{\rm QCD}\left(\tau,s_0\right)/{\cal L}_0^{\rm QCD}\left(\tau,s_0\right)$
for hybrid bottomonium is shown  as a function of the Borel scale $1/\tau$ for
the optimized value $s_0=140\,{\rm GeV^2}$ (solid curve).  
For comparison the ratio is also shown for $s_0=155\,{\rm GeV^2}$ (upper dotted curve) and $s_0=116\,{\rm GeV^2}$ (lower dotted curve). 
The uppermost dashed curve represents the $s_0\to\infty$ limit corresponding to the bound $M<10.84\,{\rm GeV}$.
Central values of the QCD parameters have been used.
}
\label{bottom_opt}
\end{figure}

The details of the bottomonium hybrid analysis proceeds in a very similar fashion as the charmonium hybrid case.  The sum-rule window of validity for the optimized $s_0=140\,{\rm GeV^2}$ is shown in Fig.~\ref{bottom_window}, and using the same methodology described above, we establish the region  $s_0>115\,{\rm GeV^2}$
and $2.9\, {\rm GeV}<1/\sqrt{\tau}<4.2\,{\rm GeV}$  for locating  an optimized prediction.  The optimization procedure described above yields  $s_0=140\,{\rm GeV^2}$ and  the predicted bottomonium hybrid mass $10.64\,{\rm GeV}$ shown in Fig.~\ref{bottom_opt}  in addition to larger and smaller $s_0$ values, including the $s_0\to\infty$ limit used for obtaining mass bounds.  The bottom quark mass and $\langle g^3G^3\rangle$ variations (\ref{mb_mass},\ref{GGG_value}) still dominate uncertainties resulting in the final mass prediction $M=(10.64\pm 0.19)\,{\rm GeV}$. Once again, the basic scales of our predictions are in good agreement with the $1^{--}$ results \cite{Qiao:2010zh}.

\section{Conclusions}
\label{theConclusion}
In this paper we have calculated the leading order perturbative, $\langle \alpha G^2\rangle$, and $\langle g^3G^3\rangle$ contributions to the pseudoscalar ($J^{PC}=0^{-+}$) heavy quark hybrid correlation function.  A full calculation of the  perturbative and $\langle \alpha G^2\rangle$ terms has been performed, and the imaginary parts confirm the results of Refs.~\cite{Govaerts:1985fx,Govaerts:1984hc}.  However, the $\langle g^3G^3\rangle$ contributions have not previously been calculated, and  the full contribution to  the correlation function was needed because  the imaginary part was not sufficient to determine the Laplace sum-rules.  

In the absence of the $\langle g^3G^3\rangle$ contributions, a stable Laplace sum-rule prediction of the $0^{-+}$ charmonium and bottomonium hybrids was not achieved even with sophisticated coupled sum-rule methods  \cite{Govaerts:1984hc,Govaerts:1986pp}.  However, the $\langle g^3G^3\rangle$ effects are able to stabilize the Laplace sum-rule mass analysis and result in the predictions $M=(3.82\pm 0.13)\,{\rm GeV}$ for the charmonium hybrid and  $M=(10.64\pm 0.19)\,{\rm GeV}$ for hybrid bottomonium.  The uncertainties in our mass predictions only include effects of the QCD input parameters;  we make no attempt to estimate the effect of higher-loop or other theoretical uncertainties.  We emphasize that the $\langle g^3G^3\rangle$ uncertainty is clearly observable, demonstrating that the dimension-six contributions are significant enough to stabilize the analysis. 
Ref.~\cite{Qiao:2010zh} previously found  a similar stabilizing effect of the $\langle g^3G^3\rangle$ contributions for mass predictions of  $1^{--}$ hybrid charmonium and bottomonium, and the sum-rule scales of the $1^{--}$ and $0^{-+}$ systems are in qualitative agreement.

The results of our analysis may have implications concerning the Y(3940) first observed by the Belle
Collaboration~\cite{Choi:2005} and seemingly verified by the BaBar Collaboration~\cite{Aubert:2007vj} although at 
the significantly lower mass of 3915~MeV.
There seems to be an emerging consensus that the Y(3940) and the X(3915)~\cite{Uehara:2009tx} 
are the same particle whereas the Y(3940) and the X(3940)~\cite{Abe} are 
distinct~\cite{delAmoSanchez:2010jr,Aushev:2010zza,Aubert:2007rva}.
In what follows, we adopt this point of view.

As first noted in the paper announcing its discovery~\cite{Choi:2005}, 
the Y(3940) is a legitimate charmonium hybrid candidate.
It is observed in B decays which, as argued in~\cite{Close:1997wp}, are thought to be prime charmonium
hybrid hunting grounds.
Also, to date, the only hadronic decay mode detected is to $\omega\,J/\psi$~\cite{Aushev:2010zza,Aubert:2007rva},
an observation difficult to reconcile with a conventional charmonium meson 
assignment considering the kinematically-allowed $D\overline{D}$ and $D\overline{D}^{*}$ channels.
Such peculiar decay signatures are, however, consistent with a hybrid interpretation 
as there exists a flux tube model-inspired selection rule 
which heavily suppresses hybrid decays to pairs of S-wave mesons~\cite{Page:1996rj,Close:1994hc,Isgur:1985vy}.

The $\omega\,J/\psi$ decay mode allows for a straightforward identification of the Y(3940)
as an isosinglet with $C=+$. Unfortunately, the  $J^P$ assignment is not so simple;
additional effort is required to identify the needed spin and parity quantum numbers.

Optimistically assuming that the Y(3940) will eventually be identified as a pseudoscalar
state, comparing our mass prediction of  3820~MeV (and 130~MeV uncertainty) to the measured value of 3915~MeV
provides additional evidence in favour of an interpretation of the Y(3940) as a charmonium hybrid or
at least as a resonance admitting a significant hybrid component.

\bigskip
\noindent
{\bf Acknowledgements:}  We  are grateful for financial support
from the Natural Sciences and Engineering Research Council of Canada (NSERC). We thank John Gracey for generously providing his advice on loop calculations.


\end{document}